\documentclass[aps,twocolumn,showpacs,floatfix]{revtex4-1}
\usepackage[utf8]{inputenc}
\usepackage{amsmath}
\usepackage{amsfonts}
\usepackage{amssymb}
\usepackage{subfigure}
\usepackage{graphicx}
\usepackage[breaklinks=true,colorlinks]{hyperref}

\newcommand{\eg}{\emph{e.g.}\ }
\newcommand{\ie}{\emph{i.e.}\ }
\newcommand{\cf}{cf.\ }

\newcommand{\ii}{\mathrm{i}}

\newcommand{\cov}{\operatorname{Cov}}
\newcommand{\ket}[1]{\left|{#1}\right\rangle}

\newcommand{\ew}[1]{\left\langle{#1}\right\rangle}

\newcommand{\kB}{k_\mathrm{B}}
\newcommand{\DC}{\Delta_\mathrm{c}}
\newcommand{\omrec}{\omega_\mathrm{R}}
\newcommand{\Erec}{E_\mathrm{R}}
\newcommand{\ac}{a_\mathrm{c}}
\newcommand{\as}{a_\mathrm{s}}
\newcommand{\Cp}{\mathcal{C}_p}
\newcommand{\B}{\mathcal{B}}

\begin{document}

\title{Subrecoil cavity cooling towards degeneracy: A numerical study}

\author{Raimar M. Sandner}
\affiliation{Institut f{\"u}r Theoretische Physik, Universit{\"a}t Innsbruck, Technikerstra{\ss}e~25, 6020~Innsbruck, Austria}

\author{Wolfgang Niedenzu}
\affiliation{Institut f{\"u}r Theoretische Physik, Universit{\"a}t Innsbruck, Technikerstra{\ss}e~25, 6020~Innsbruck, Austria}

\author{Helmut Ritsch}
\email{Helmut.Ritsch@uibk.ac.at}
\affiliation{Institut f{\"u}r Theoretische Physik, Universit{\"a}t Innsbruck, Technikerstra{\ss}e~25, 6020~Innsbruck, Austria}

\date{August 14, 2013}

\pacs{37.30.+i, 37.10.-x, 37.10.Vz}

\begin{abstract}
We present a detailed numerical analysis of the temperature limit and timescale of cavity cooling of a dilute gas in the quantum regime for particles and light. For a cavity with a linewidth smaller than the recoil frequency efficient cooling towards quantum degeneracy is facilitated by applying a tailored sequence of laser pulses transferring the particles towards lower momenta. Two-particle Monte Carlo wave function simulations reveal strongly improved cooling properties for a ring versus a standing-wave geometry. Distinct quantum correlations and cooling limits for bosons and fermions demonstrate quantum statistical effects. In particular, in ring cavities the photon-mediated long-range interaction favours momentum-space pairing of bosons, while fermion pairs exhibit anti-correlated or uncorrelated momenta. The results are consistent with recent experiments and give encouraging prospects to achieve sufficient conditions for the condensation of a wide class of polarisable particles via cavity cooling.
\end{abstract}

\maketitle

\section{Introduction}
Cavity cooling has been proposed and successfully demonstrated in practise as a quite general method to cool polarisable point particles already more than a decade ago~\cite{Horak1997,Domokos2001,Schleier2011,Reiserer2013}. Several important experimental demonstrations using various systems and geometries were recently achieved~\cite{Ritsch2013}, also in close connection to optomechanical cooling~\cite{Schulze2010,Brahms2012}. In its generic form, this cooling mechanism works without the need of resonant excitation and spontaneous emission and reveals a final temperature only limited by the linewidth $\kappa$ of the cavity mode involved, \ie $\kB T \approx \hbar \kappa$~\cite{Ritsch2013}. For deeply trapped particles one can operate in the sideband cooling regime with trapping frequency $\nu$ larger than $\kappa$, where ground-state cooling can be expected~\cite{Zippilli2005,Schulze2010}. In the opposite case, where the cavity-generated optical potential is shallow compared to the particles' kinetic energy, the situation is much more complex as we have to consider the full sinusoidal dependence of the mode function. 

As the optical potential is periodic with half of the wavelength, photon scattering only exchanges multiples of two recoil momenta between the atomic motion and the cavity field. Microscopically, this corresponds to a consecutive absorption and stimulated emission process of cavity photons. Hence, on the one hand one could argue that cavity cooling cannot achieve temperatures below the recoil limit due to the half-wavelength periodicity of the cavity potential~\cite{stecher1997all}. On the other hand, a cavity with frequency and energy resolution below a single recoil could be suited to surpass this limit by energy selection. Strong evidence for this effect has been reported in recent experiments in Hamburg, where cavity cooling on the subrecoil scale has been observed~\cite{Wolke2012}.

\par

In this letter we study the quantum limit of cavity cooling in great detail by help of quantum Monte Carlo wave function simulations~\cite{Dum1992,Molmer1993} of the corresponding master equation. In order to allow investigations of particle quantum statistics and correlations, we need to consider at least two particles coupled to the cavity field, which itself can support a single or two modes. Even when confining the particle motion along the cavity axis we still have to account for at least three independent quantum degrees of freedom. Despite restriction to only a few photons per mode and a not too high spatial resolution of the wave function, the associated Hilbert space typically consists of several ten thousand dimensions, which requires substantial computational effort. Here we make use of C++QED~\cite{Vukics2012}, a freely available framework for simulating open quantum dynamics.

In the following, upon introducing our system Hamiltonian and master equation and defining the operating parameters we first present the results of a three-stage cooling process from a thermal distribution towards the quantum ground state. We then study the mean kinetic energy and address the influence of invariant subspaces on the final temperature and state. In the last part we analyse the appearance of momentum correlations during the cooling process, which can be nicely seen and interpreted via the two-particle momentum distributions.

\section{Theoretical model}

\begin{figure}
 \centering
 \includegraphics[width=0.8\columnwidth]{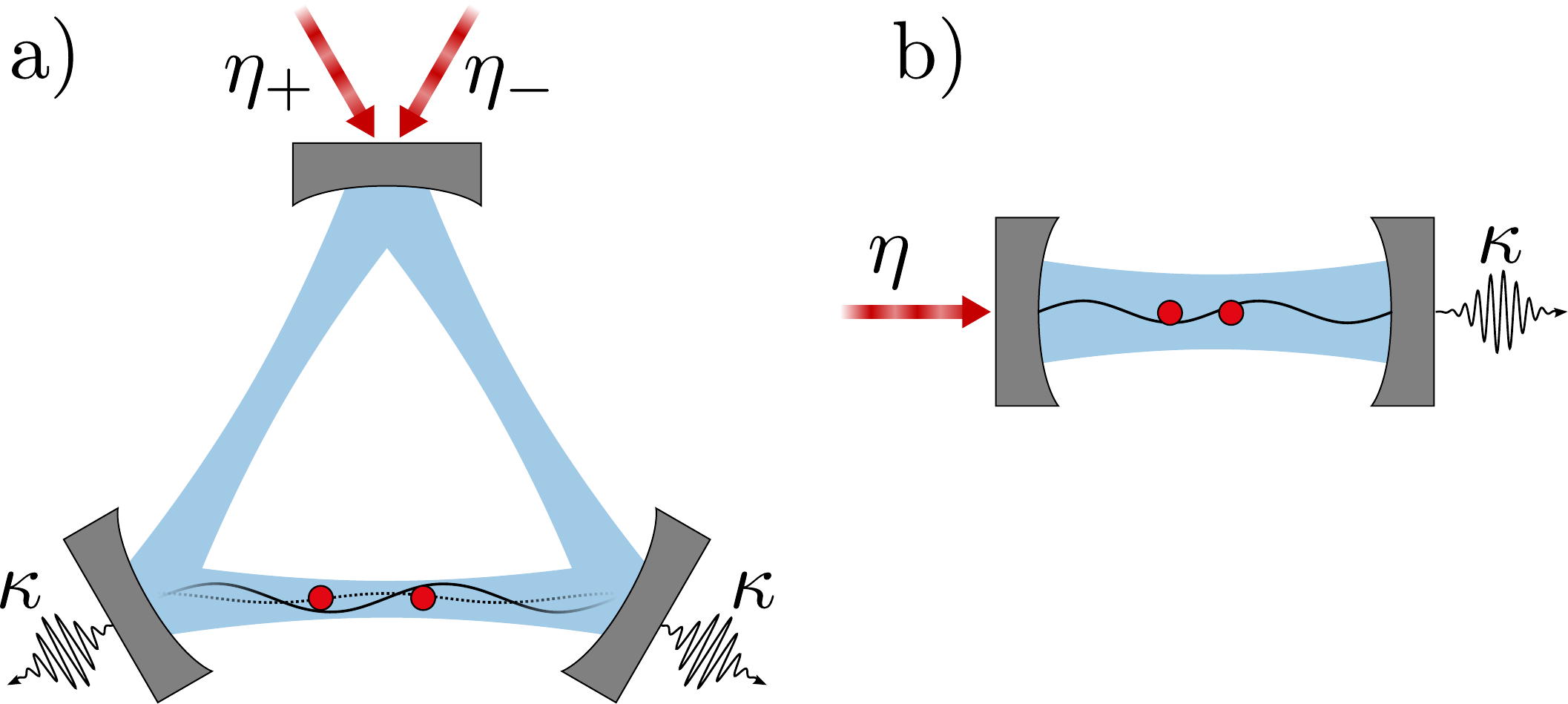}
 \caption{(Colour on-line) Two identical particles interacting with a) two degenerate modes of a ring resonator and b) a single standing-wave mode.}\label{fig_system}
\end{figure}

We consider two polarisable particles (\eg two-level atoms, molecules, nanobeads) of mass $m$ moving along the axis of an optical resonator, as depicted in fig.~\ref{fig_system}. 

Assuming large detuning between the light and optical particle resonances, the particles' internal excitation will remain small and spontaneous emission can be neglected~\cite{Ritsch2013}. The intracavity light is described by one or two degenerate standing-wave eigenmodes. For our studies the description of the two-mode ring resonator (\cf fig.~\ref{fig_system}a) in terms of standing-wave modes $\propto \sin (kx)$ and $\propto \cos (kx)$ with wave number $k$~\cite{Schulze2010,Niedenzu2010} is more convenient than the mathematically equivalent description with a set of running-wave eigenmodes $\propto \exp(\pm \ii kx)$. It allows easier comparison with the linear resonator (\cf fig.~\ref{fig_system}b) by formally restricting one of the modes to zero photons. The coherent interaction of particles and field modes of the ring resonator is then described by the Hamilton operator~\cite{Schulze2010}
\begin{multline}\label{eq_H}
 H=\sum_{i=1}^2\left[\frac{p_i^2}{2m}+\hbar U_0\bigl(\ac^\dagger\ac\cos^2(kx_i)+\as^\dagger \as\sin^2(kx_i)\bigr)\right]\\
 +\frac{\hbar U_0}{2}\left(\ac^\dagger\as+\as^\dagger\ac\right)\sum_{i=1}^2\sin(2kx_i)\\
 -\hbar\DC \ac^\dagger \ac - \hbar\DC\as^\dagger\as-\ii\hbar\eta\left(\as-\as^\dagger\right).
\end{multline}
Here $x_i$ and $p_i$ are the particles' centre-of-mass position and momentum operators, respectively, $U_0<0$ is the potential depth per photon and $\as^\dagger$ ($\ac^\dagger$) is the creation operator for sine (cosine) mode photons. The pump laser of effective strength $\eta$ is detuned by $\DC:=\omega_\mathrm{p}-\omega_\mathrm{c}$ from the bare cavity resonance frequency. Note that an effective driving of the sine mode (as considered here) can be realised by symmetrically driving the two counter-propagating running-wave modes with an appropriate phase relation~\cite{Schulze2010}. Whilst the first line of the Hamiltonian~\eqref{eq_H} accounts for the two potentials created by the degenerate light modes, the second line describes the particle-mediated coherent redistribution of photons between the two modes, which is responsible for photon scattering to the unpumped cosine mode.
\par
The Hamiltonian evolution, however, cannot take track of the dissipative process of photon leakage through the resonator's mirrors. Indeed, the system needs to be characterised by its density operator $\rho$ with time evolution
\begin{equation}\label{eq_master}
\dot\rho=\frac{1}{\ii\hbar}\left[H,\rho\right]+\mathcal{L}_\mathrm{c}\rho+\mathcal{L}_\mathrm{s}\rho,
\end{equation} where the Liouvillean superoperators in this master equation explicitly read $\mathcal L_i\rho=\kappa\left(2a_i\rho a_i^\dagger - a_i^\dagger a_i\rho - \rho a_i^\dagger a_i \right)$ for $i\in\{\mathrm{c},\mathrm{s}\}$~\cite{Gardiner2000} with the photon number decay rate $2\kappa$. Note that the linear cavity (\cf fig.~\ref{fig_system}b) is obtained from eq.~\eqref{eq_master} by formally setting $\ac=0$.

Direct particle--particle interactions as collisions are neglected in our model. They will certainly play a role at higher densities and should be included in a next-step model. Here we concentrate on cavity cooling and only shortly come back to possible effects of collisions in the follwing section.

\section{Simulation of cavity cooling in the quantum regime}

\begin{figure*}
 \centering
 \subfigure{
   \includegraphics[width=0.8\textwidth]{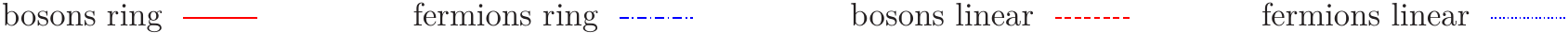}
 }
 \subfigure[\ Stage 1, $\eta=3\omrec$]{
   \includegraphics[width=0.3\textwidth]{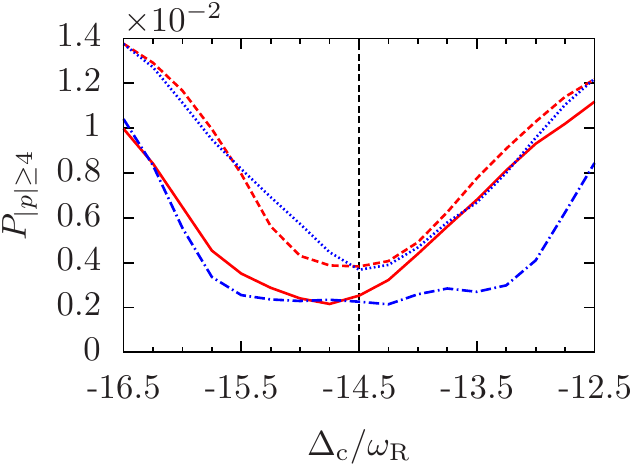}
 }
 \subfigure[\ Stage 2, $\eta=2\omrec$]{
   \includegraphics[width=0.3\textwidth]{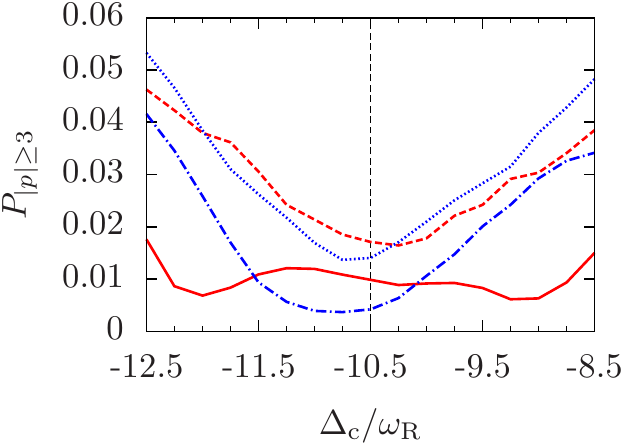}
 }
 \subfigure[\ Stage 3, $\eta=0.5\omrec$]{
   \includegraphics[width=0.3\textwidth]{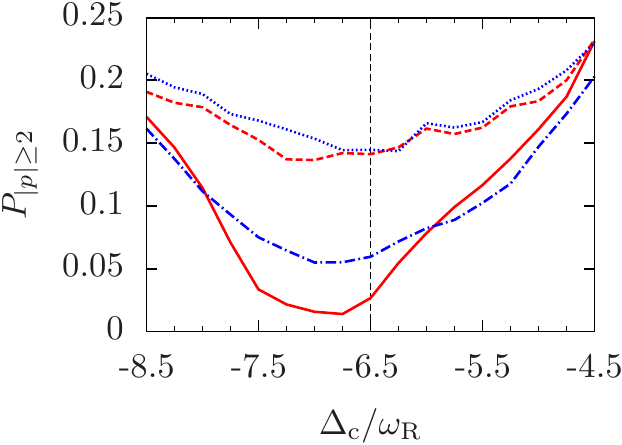}
 }
 \caption{(Colour on-line) Population fraction not addressed by the cooling process for the three cooling steps as a function of the cavity-pump detuning scanned around the free-space optimal value~\eqref{eq_eff_delta}. The final step (c) targets ground-state cooling as in the Hamburg experiment~\cite{Wolke2012}. Parameters: $U_0=-2.5\omrec$ and $\kappa=0.25\omrec$.}\label{fig_scan}
\end{figure*}

In this letter we concentrate on the weak pump case, where the intracavity light fields are responsible for friction and diffusion and mediate long-range interparticle interactions, but do not contribute to particle confinement. The latter is provided by an external magnetic or optical trap with trap frequencies much below the recoil frequency. Targeting quantum degeneracy in this trap naturally requires a full quantum treatment of both, field modes and particles, which renders a direct integration of the master equation~\eqref{eq_master} computationally unfeasible due to the large dimension of the joint Hilbert space. For this reason we run Monte Carlo wave function simulations (MCWFS)~\cite{Dum1992,Molmer1993}, where the evolution of stochastic state vectors is induced by a non-Hermitian effective Hamilton operator including damping. This deterministic propagation is interrupted by quantum jumps at random times, which can be interpreted in terms of photon counting events of the light leaving the resonator. For large ensembles the statistical mixture of such stochastic states converges towards the full solution of the master equation. The numerical implementation of these Monte Carlo wave function simulations was realised with the aforementioned C++QED framework~\cite{Vukics2012}.

Note that the chosen parameter regime is in contrast to our previous work on deeply trapped particles~\cite{Niedenzu2012}, where the strong pump field mode is used to trap the particle and can be replaced by a c-number to obtain a much smaller effective Hilbert space. 

To discretise the Hilbert space we introduce periodic boundary conditions for the particles at $kx=\pm \pi$, yielding two-particle momentum basis states $\ket{n,m}:=\ket{p_1=n\hbar k}\otimes \ket{p_2=m\hbar k}$ with integer $n,m$. We start from an empty cavity without photons and the two particles located at $kx=\pm \pi/2$ with a Gaussian distribution and zero average momentum, properly \mbox{(anti-)}symmetrised for bosons (fermions). Here we neglect the spin degrees of freedom as is justified \eg for spin-polarised particles. Other possible configurations like pairs of fermions with opposite spins, as they appear in the BCS model~\cite{Bardeen1957}, might show yet another behaviour but are beyond the scope of this letter.

In order to model the characteristics of a thermal state and to remove artefacts due to the periodic boundary conditions, we assign random phases to each basis component of the Monte Carlo trajectory initial states. For sufficiently many trajectories the initial state then becomes a statistical mixture of momentum states without any coherences. Note that here the two particles are only found in momentum states with the same parity if they are bosons or in momentum states with different parity if they are fermions. Implications of this choice will be discussed in the next section.

Standard semiclassical simulations of cavity cooling of point particles predict a final temperature $\kB T \approx \hbar \kappa $ only limited by the cavity linewidth $\kappa $, which also agrees with full quantum simulations to very low temperatures~\cite{Horak1997}. While smaller $\kappa$ leads to lower temperature, unfortunately the time scale for cooling becomes much slower as well and the effective velocity capture range is reduced~\cite{Nimmrichter2010}. This renders cooling of a wide momentum distribution towards degeneracy in a single step experimentally impractical. Hence, here we present a multi-step cooling scheme as inspired by the recent Hamburg experiment~\cite{Wolke2012}, where the operating parameters are adjusted in time to speed up cooling by first precooling the highest momenta and subsequently reducing pump power $\eta$ and detuning $|\DC|$ for the final step. 

So far we have ignored any effects stemming from direct atom--atom collisions in our effective 1D approximation of the cooling geometry. As can be seen from a Fourier transformation of the Hamiltonian~\eqref{eq_H}, momentum transfer between atomic states can only occur in portions of $\Delta p=\pm 2\hbar k$. Therefore the total population in the odd and even momentum subspaces is conserved by the Hamiltonian evolution. In a realistic experimental trap, however, there will be transverse motion and particle--particle collisions on longer timescales mixing these two families, \eg two particles with momentum $\ket{1,1}$ can scatter into the $\ket{0,2}$ state and vice versa. The fast $2\hbar k$ particle is then cavity-cooled to zero velocity, which makes this effective process unidirectional. Another important aspect of collisions has been observed in the aforementioned Hamburg experiment~\cite{Wolke2012}, where they play a crucial role in cooling particles from non-integer momentum states in the vicinity of $p=\pm2\hbar k$ to the zero momentum state. As it is not in the scope of the present work to describe these processes, but much rather the essence of cavity cooling of a thermal state with broad momentum distribution, we refrain from including collision terms in the simulations at this point.

The effective detuning of the pump is chosen such that the creation of a photon in the cavity is energetically strongly suppressed if not accompanied by momentum transfer. Cooling is achieved when a scattered photon leaves the resonator, which eliminates the re-absorption possibility~\cite{Castin1998}. In most cases this prevents standard laser cooling to reach degeneracy, although this limitation has recently been surpassed in an Innsbruck experiment~\cite{Stellmer2013}. Intuitively, in the regime $\DC<0$ the cavity-enhanced scattering to the blue sideband extracts energy from the particles' motion~\cite{Ritsch2013}. To resonantly enhance this process the detuning is chosen to match the energy difference between the final and initial atom states. From the dispersion relation for free particles with $\Delta p=-2\hbar k$ one thus obtains
\begin{equation}\label{eq_eff_delta}
 \DC-NU_0\ew{\sin^2(kx)}\stackrel{!}{=}\frac{P_\mathrm{final}^2}{2\hbar m}-\frac{P_\mathrm{init}^2}{2\hbar m}=4\omrec\left(1-\frac{K_\mathrm{init}}{k}\right),
\end{equation}
where $N=2$ is the number of particles, $P=\hbar K$, $K_\mathrm{init}\geq2k$ and $\omrec\equiv\Erec/\hbar:=\hbar k^2/(2m)$ is the recoil frequency. As the particles are nearly homogeneously distributed in position space, the bunching parameter $\B:=\ew{\sin^2(kx)}$ determining the cavity frequency shift induced by the particles is well approximated by $\B\approx\frac{1}{2}$. In the regime of weak cavity pump with potential depths well below the recoil energy, the particle position space distribution remains approximately homogeneous throughout the cooling process and hence the bunching parameter stays constant. For condition~\eqref{eq_eff_delta} to be relevant, a narrow cavity resonance on the recoil frequency scale is required, \ie the damping constant $\kappa$ needs to be less than the smallest possible kinetic energy transfer, $\kappa\ll4\hbar k^2/(2m)\equiv 4\omrec$. The condition of well-resolved lines also requires that the photon numbers and the coupling strength $|U_0|$ are sufficiently small to avoid broadening of the resonance. 

Because of the quadratic dispersion relation and the small linewidth only one process with momentum transfer $|p|=n\hbar k\rightarrow |p|=(n-2)\hbar k$ with integer $n\ge 2$ can be resonant for a given detuning. In order to efficiently cool an initial state which contains momentum contributions up to $|p|=4\hbar k$, in analogy to~\cite{Wolke2012}, we thus apply a pulse sequence with three separate stages, which aim to transfer $|p|=4\hbar k\rightarrow |p|=2\hbar k$, $3\hbar k\rightarrow\hbar k$ and $2\hbar k\rightarrow0\hbar k$, respectively. Ideally, the final state then only has contributions of the two lowest momentum states with $p=0\hbar k$ and $p=\pm\hbar k$, with their relative magnitude determined by the initial distribution in the two (odd and even) momentum subspaces. In an experiment one could of course also envisage a continuous frequency ramp of the pump laser or cavity resonance. 

In order to determine the optimal detuning for the three-step cooling sequence, we scan the resonance around the anticipated optimal detuning as given in eq.~\eqref{eq_eff_delta}. In fig.~\ref{fig_scan} we plot the residual population remaining in states with momenta larger or equal to the momentum state which is targeted for depletion as a function of the detuning and choose the minimum as the detuning for this stage. Due to photons being generated during cooling the particle energies are slightly shifted and also the bunching parameter can change by a small amount. Thus the optimum detuning deviates from the expected free-space value~\eqref{eq_eff_delta}, which could also be attributed to an effective mass of the particles in the weak cavity-generated lattice. 

To study quantum-statistical effects explicitly we compare both cases of bosons and fermions. Interestingly, pronounced differences in cooling rates and pairing appear already in the first cooling stage applied to a relatively hot ensemble not too close to degeneracy. In order to investigate the influence of cavity-mediated interactions in greater detail we compare a ring cavity geometry (\cf fig.~\ref{fig_system}a) and a linear cavity as depicted in fig.~\ref{fig_system}b, \ie by formally setting~$\ac=0$ in the Hamiltonian~\eqref{eq_H}. Here the differences are clearly due to the interactions generated by the second unpumped cavity mode. 

\section{Ground-state cooling}

\begin{figure}
 \centering
 \includegraphics[width=0.9\columnwidth]{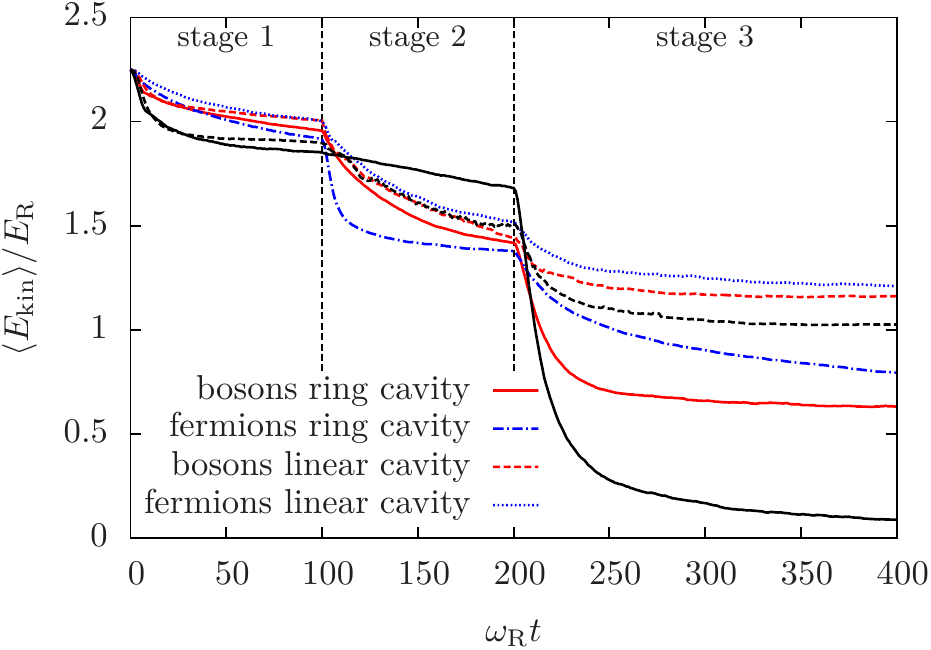}
 \caption{(Colour on-line) Time evolution of the single particle kinetic energy in an optimised three-stage process. Detunings for the three stages: $\DC/\omrec=(-14.75,-12,-7)$ (ring cavity bosons), $\DC/\omrec=(-14.5,-10.25,-7)$ (linear cavity bosons), $\DC/\omrec=(-14.25,-10.75,-6.75)$ (ring cavity fermions) and $\DC/\omrec=(-14.5,-10.75,-6.25)$ (linear cavity fermions). The other parameters are the same as in fig.~\ref{fig_scan}. Black lines: initial state containing only even momenta for bosons in a ring (solid) and a linear cavity (dotted).}\label{fig_cooling}
\end{figure}

\begin{figure}
 \centering
 \includegraphics[width=0.8\columnwidth]{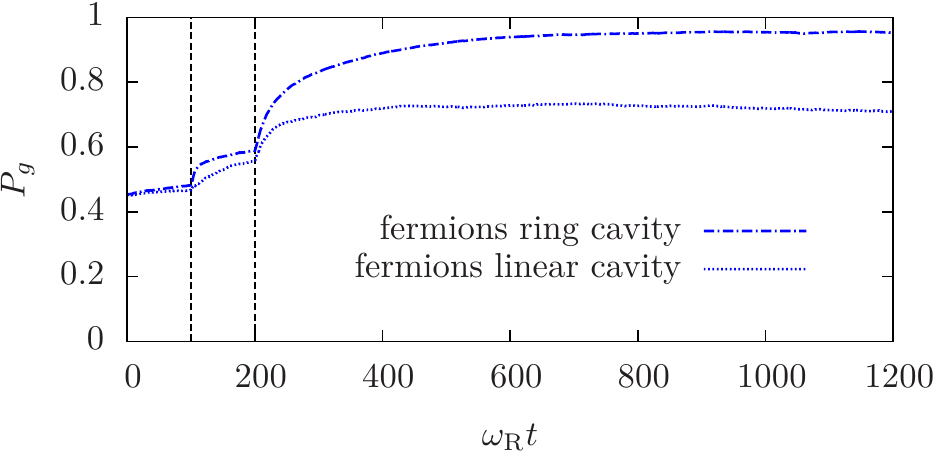}
 \caption{(Colour on-line) Projection onto the fermionic ground state $P_\mathrm{g}$. The parameters are the same as in fig.~\ref{fig_cooling}.}
 \label{fig:Pg}
\end{figure}

We now examine the results of the MCWFS for the optimised three-step cooling sequence, averaged over a sufficiently large ensemble of 1000 trajectories. Figure~\ref{fig_cooling} shows the mean kinetic energy per particle $\ew{E_{\text{kin}}}$ as a function of time. Note that for our choice of the initial state (\cf previous section) half of the population resides in odd and even momentum states, respectively. Hence the best possible cooling achieves $\ew{E_{\text{kin}}}=0.5 E_\mathrm{R}$. For bosons, where in our initial state both particle momenta always have the same parity, this means that only half of the population can be cooled to the ground state $\ket{0,0}$. For fermions, however, the ground states are the anti-symmetrisations of $\ket{\pm 1,0}$. As these states have mixed momentum parity, just like the initial state for fermions, it is in principle possible to cool the ensemble to quantum degeneracy as seen in fig.~\ref{fig:Pg}. The black lines in fig.~\ref{fig_cooling} correspond to bosons with an initial condition where only even momentum states are populated. In the ring cavity case the mean kinetic energy approaches zero and quantum degeneracy is reached. In all cases cooling in the linear cavity is not as efficient compared to the ring cavity geometry. The reason for this will be discussed below.
\par

\begin{figure}
 \centering
 \includegraphics[width=0.8\columnwidth]{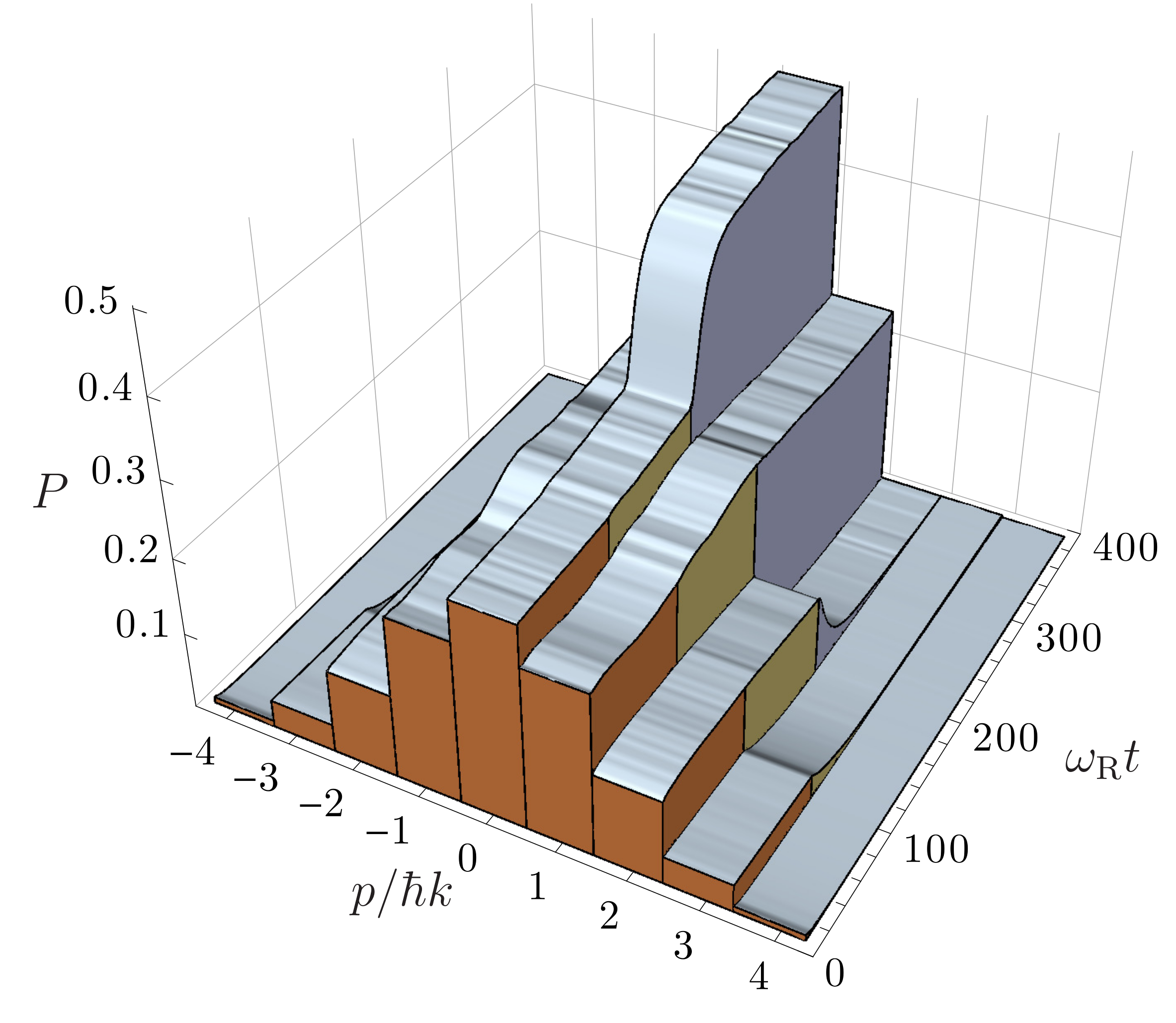}
 \caption{(Colour on-line) Single-particle momentum distribution for bosons in a ring cavity. The initially broad distribution is cavity-cooled to $|p|=0,\pm1 \hbar k$ momentum states. The parameters are the same as in fig.~\ref{fig_cooling}.}\label{fig_cooling_hist}
\end{figure}

\begin{figure}[tbp]
 \centering
 \includegraphics[width=0.85\columnwidth]{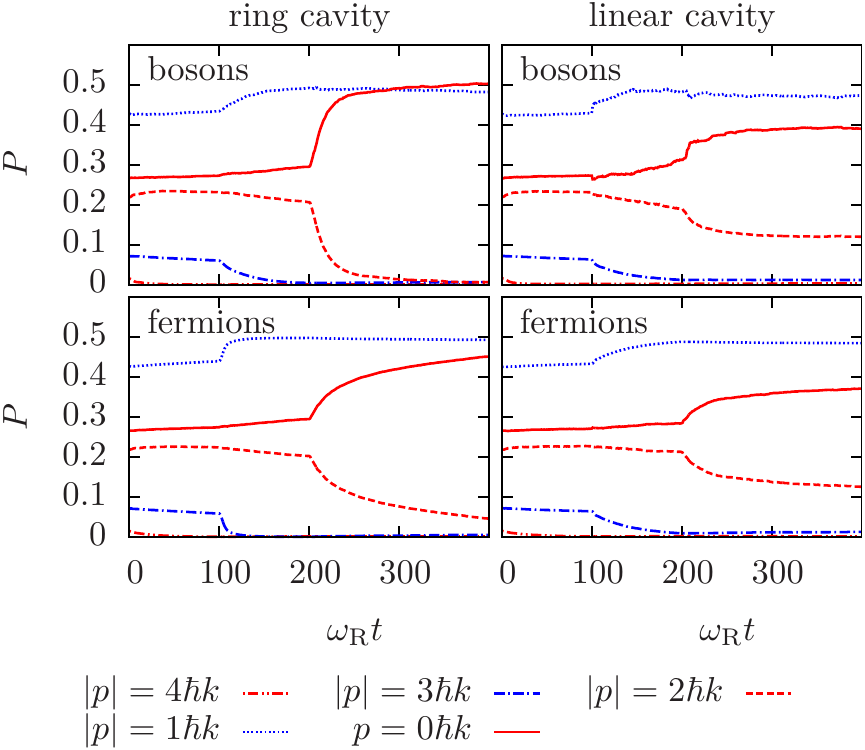}
 \caption{(Colour on-line) Occupation probabilities of the lowest few momentum eigenstates for bosons in a ring cavity (upper left), bosons in a linear cavity (upper right), fermions in a ring cavity (lower left) and fermions in a linear cavity (lower right). Same parameters as in fig.~\ref{fig_cooling}.}
 \label{fig:populations}
\end{figure}

In figs.~\ref{fig_cooling_hist} and~\ref{fig:populations}~the single-particle momentum distribution is plotted over time. It can be seen that each stage redistributes states with a distinct momentum $|p|=n\hbar k$ to corresponding lower momentum states with $|p|=(n-2)\hbar k$. The greater efficiency of a ring cavity compared to a linear cavity is most apparent in the last stage, where in the standing-wave case a considerable fraction of the population remains trapped in the $|p|=2\hbar k$ states. Figure \ref{fig:populations} shows that for a ring cavity only a very tiny amount of population is left in the $|p|=2\hbar k$ momentum state. Its ratio to the $p=0$ population can be used as a first estimate for an effective temperature in the even momentum state space via $P_2/P_0 \approx \exp(-4\Erec/\kB T_\mathrm{eff})$, which hints for a subrecoil cooling behaviour. 

\par
\par
\par

\section{Qualitative difference between ring cavity and linear cavity}

The fact that not all of the population is transferred to $|p|=0,\pm1\hbar k$ momentum in a standing wave is apparent from the final momentum space distributions. This effect is connected to subspaces of the two-particle Hilbert space which are invariant under the Hamiltonian $\eqref{eq_H}$ for $\ac=0$ and do not include the lowest energy states. Any population initially present in one of these subspaces is thus inaccessible to ground-state cooling. A typical example is the bosonic state
\begin{equation}
 \ket{\psi}=\bigl(\ket{0,2}+\ket{2,0}-\ket{0,-2}-\ket{-2,0}\bigr)/2.
 \label{eq:dark}
\end{equation}
This state is a dark state for the linear cavity and the particles will not scatter any photons. On the other hand, in a ring cavity the interference term between the two modes (\ie the second line of eq.~\eqref{eq_H}) breaks this symmetry and the atoms redistribute light between the two modes while being transferred to the bosonic ground state $\ket{0,0}$. As mentioned above, collisions between particles or an external atom trap decrease the lifetime of such a dark state preventing population accumulation. This might explain the absence of these trapping states in the experiment~\cite{Wolke2012}.

\section{Momentum correlations}

Let us finally look at the connection between particle correlations, cavity cooling and quantum statistics in our simulations. Already in the classical limit particle correlations lead to important modifications of single-mode cavity cooling~\cite{Asboth2004}. Classical simulations of ring cavities reveal a strong damping of the centre-of-mass motion leading to momentum space (anti-)pairing of particles~\cite{Gangl2000}. On the contrary, for the case of two deeply trapped quantum particles or membranes in a ring resonator positive momentum correlations and even entanglement emerged~\cite{Niedenzu2012}. The former can be quantified by the momentum correlation coefficient (normalised covariance)
\begin{equation}\label{eq_Cp}
 \Cp:=\frac{\cov(p_1,p_2)}{\Delta p_1\Delta p_2}=\frac{\ew{p_1p_2}-\ew{p_1}\ew{p_2}}{\Delta p_1\Delta p_2}.
\end{equation}

\begin{figure}[tbp]
 \centering
 \includegraphics[width=0.8\columnwidth]{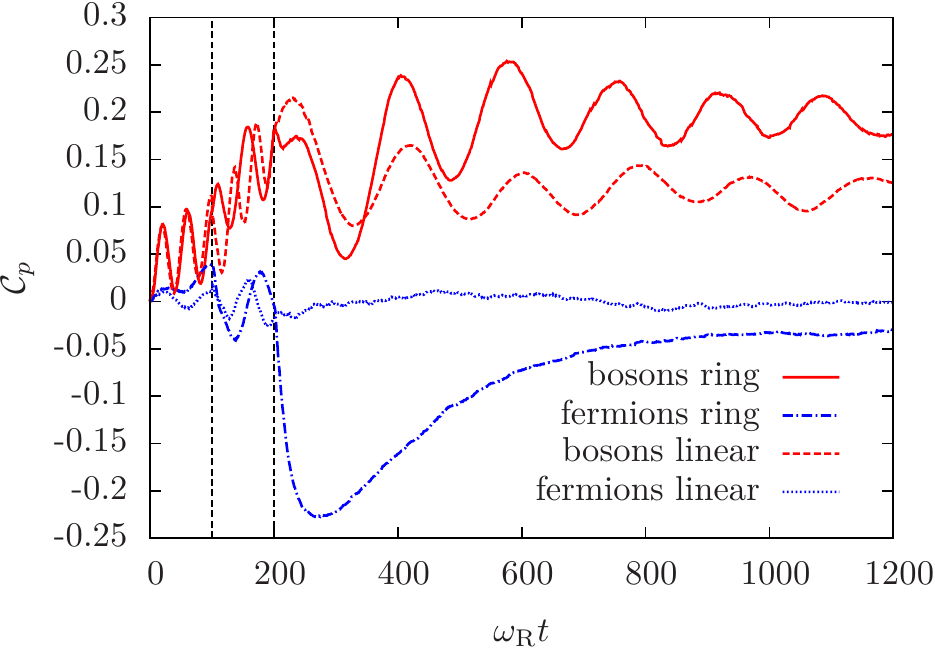}
 \caption{(Colour on-line) Momentum correlation coefficient~\eqref{eq_Cp} for different particle statistics and cavity geometries. The parameters are the same as in fig.~\ref{fig_cooling}.}
 \label{fig:momcorr}
\end{figure}

\begin{figure}
 \subfigure[\ Bosons ring, $\omrec t=600$]{\includegraphics[width=0.48\columnwidth]{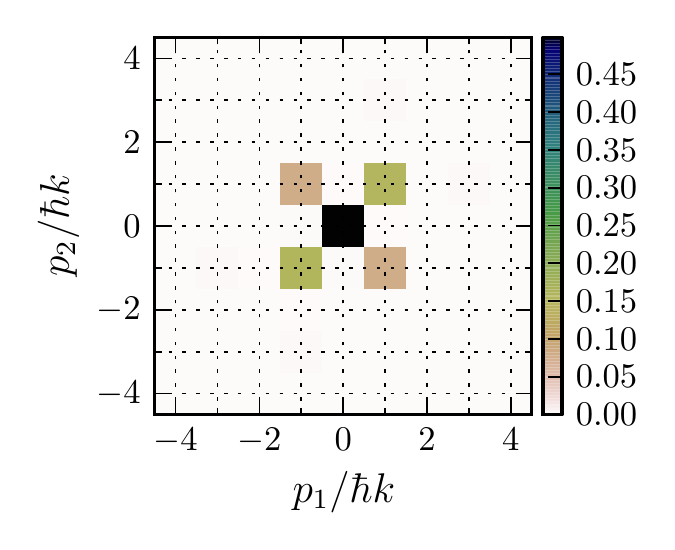}}
 \subfigure[\ Fermions ring, $\omrec t=300$]{\includegraphics[width=0.48\columnwidth]{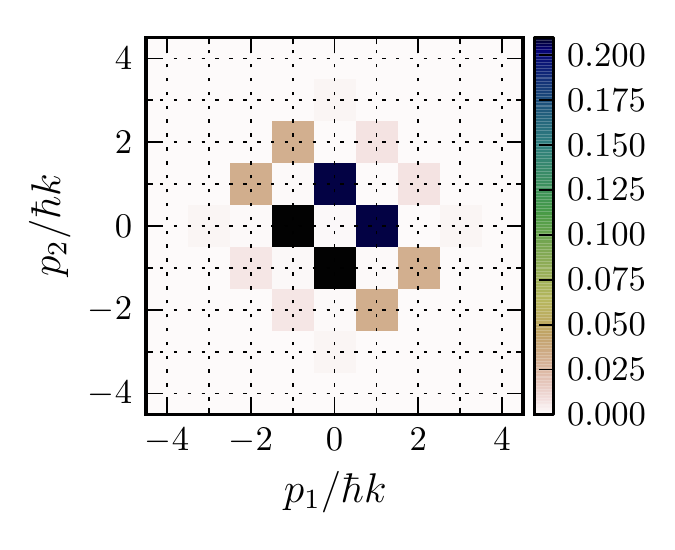}}
 \caption{(Colour on-line) Two-particle momentum population at times where the (anti-)correlations are maximal according to fig.~\ref{fig:momcorr}. Positive correlations (bosons) and negative correlations (fermions) can be seen in the occupation profiles being elongated along different diagonals.}
 \label{fig:2Dmomentum}
\end{figure}

Here we extend this study to indistinguishable particles obeying either Bose or Fermi statistics. In fig.~\ref{fig:momcorr} we show $\Cp$ as a function of time exhibiting a complex evolution. Note that in the unperturbed ground state bosons as well as fermions are uncorrelated. Thus all population in the lowest energy states does not contribute to momentum correlations and ideal cooling should eliminate all correlations. The small excited state fraction, however, reveals strong correlations as highlighted in the two-particle momentum space distribution shown in fig.~\ref{fig:2Dmomentum}. As one might expect, fermions are anticorrelated and bosons are correlated in momentum space. This behaviour is substantially more pronounced in ring cavities and, in the case of bosons, survives very long cooling times, even exhibiting an oscillatory behaviour. Note that positive pair correlations with zero average momentum strongly point towards the appearance of momentum-entangled states.

\par

Let us remark here that these correlations can be much stronger if the pump is tuned away from optimal cooling. One can enhance certain correlation by targeting the corresponding states instead of the ground state in a tailored pulse sequence. A detailed analysis, however, would go beyond the scope of this work. 

\section{Conclusion and outlook}

Extensive numerical simulations confirm the experimental observation of subrecoil cooling towards degeneracy using high-$Q$ cavities. We show that efficient cooling of a thermal distribution requires either collisional redistribution or several cavity modes to avoid the appearance of decoupled momentum subspaces, which cannot be cooled solely by a single mode as in the recent Hamburg experiment~\cite{Wolke2012}. We predict that bosons and fermions can be cooled in the same way towards a high occupation of their respective ground states. 

As a ring cavity works generally much better than a single-mode resonator one can expect further improvements including more modes similar to the classical case~\cite{Domokos2002}. In principle, a linear cavity supports many non-degenerate modes separated by the free spectral range, so that one could also envisage an extension of the studied scheme to such a multimode setup. The sequential pulses can then be replaced by differently detuned driving lasers for these modes. This would allow to address the three cooling stages simultaneously and hence to reduce and optimise the cooling time. Such a setup also allows to generate and observe more exotic particle--particle correlations.

As a bottom line we suggest that cavity cooling is a viable route towards producing a degenerate gas of a wide class of polarisable particles without the need of evaporation. Hence it could be applied at much lower densities and particle numbers once a sufficient number can be loaded inside a high-$Q$ resonator.

\begin{acknowledgments}
This work has been supported by the EU-ITN CCQED project, by the Austrian Science Fund FWF through project F4013 and by the Austrian Ministry of Science BMWF as part of the UniInfrastrukturprogramm of the Focal Point Scientific Computing at the University of Innsbruck.
\end{acknowledgments}
\end{document}